 \documentclass [12pt,a4paper      ]{article}
\usepackage{times}

\DeclareFontFamily{OT1}{times}{}
\DeclareFontShape {OT1}{times}{m }{n }{ <-> ptmr }{}
\DeclareFontShape {OT1}{times}{bx}{n }{ <-> ptmb }{}
\DeclareFontShape {OT1}{times}{m }{it}{ <-> ptmri}{}
\DeclareFontShape {OT1}{times}{bx}{it}{ <-> ptmbi}{}
\usepackage{amsmath}
\usepackage{amsfonts}
\usepackage{amssymb}
\usepackage{latexsym}
\newcommand{\cl}{C \kern -0.1em \ell} 

\newcommand{\CON}{\overline}          

\newcommand{\Scal}{\mathbb{S}}        

\newcommand{\VEC}{\vec{\kern +.1em[}} 
\newcommand{\TOR}{\vec{\kern +.2em]}} 
\newcommand{\BRA}{\langle\kern -.2em\langle} 
\newcommand{\KET}{\rangle\kern -.2em\rangle} 

\newcommand{\DUA}{\widetilde}         

\numberwithin{equation}{section}
\begin{document}

\title{\bf\vspace{-2.5cm} The Tensor Analytical Relationships of
                          Dirac's Equation\footnote{\emph{Editorial note}:
                          Published in
                          Zeits.\ f.\ Phys.\ {\bf 57} (1929) 447--473,
                          reprinted and translated in~\cite{LAN1929B}.
                          This is Nb.~1 in a series of four papers
                          on relativistic quantum mechanics
                          \cite{LAN1929B,LAN1929C,LAN1929D,LAN1930A}
                          which are extensively discussed
                          in a commentary by Andre Gsponer
                          and Jean-Pierre Hurni \cite{GSPON1998B}.
                          Initial translation by J\'osef Illy and 
                          Judith Konst\'ag Mask\'o. Final translation
                          and editorial notes by Andre Gsponer.}}

\author{By Cornel Lanczos in Berlin\\ (Received on July 17, 1929)}

\date{Version ISRI-04-10.5 ~~ \today}

\maketitle

\begin{abstract}

Dirac's equation of the electron will be discussed by using quaternions as the basis of a new formalism which seems to be very well adapted to the problem. The transformation properties of the equations as well as the invariant and covariant [bilinear] constructions of Dirac's theory are developed uniformly and systematically. A method of obtaining a covariant formulation of the equations using customary tensor calculus also offers itself unequivocally if we duplicate the Dirac equations. The resulting system consists of two parts. The first part corresponds to the coupling of an antisymmetric tensor with a vector of the same structure as in the relationship between electromagnetic field strength and current vector in Maxwell's equations. The other part yields a ``feedback'' in the form of a reaction of the current vector upon the field strength of the same structure as in the case of the well-known connection between the vector potential and the field strength for the electromagnetic field.

\end{abstract}

\section{Introduction}

Dirac's equation is based on two viewpoints. On the one hand, it should be a first-order linear differential equation and, on the other hand, the iterated application of the equation (for the field-free case) should yield the Schr\"odinger wave equation.\footnote{\emph{Editorial Note}: By Schr\"odinger wave equation Lanczos means the \emph{relativistic} Schr\"odinger wave equation which today is usually referred to as the Klein-Gordon equation.}  The latter aspect in practice implies the invariance under Lorentz transformations without stating that the transformations involved should necessarily have a common vector analytical meaning.  The matrix calculus and the operator method have the advantage that they are able to provide a solution to the problem without being concerned with the requirements of normal tensor analysis. This advantage is diminished by the uncomfortable fact that in a physical theory one has to work with quantities which do not allow interpretation by the normal concepts of tensor analysis --- though these concepts have otherwise proved extremely useful in describing natural phenomena. On these grounds, several trials were made to avoid the operator method and to bring the Dirac equation into a form which could be interpreted in terms of normal vector analytical concepts. C.\ G.\ Darwin\footnote{Proc.\ Roy.\ Soc.\ {\bf 120}, 621, 1928; Nature {\bf 123}, 203, 1929.\label{Darwin}} pointed out an analogy with Maxwell's equations in the special case when the mass term of Dirac's equation vanishes. On the other hand, Madelung\footnote{Zeits.\ f.\ Phys.\ {\bf 54}, 303, 1929.} worked out a system of equations which can be considered to be a generalization of Maxwell's equations and which enables the Dirac equation to be obtained in a natural interpretation.  For this system, the invariance under Lorentz transformations remains a problem.\footnote{\emph{Remark during revision:} I was informed from a kind letter by Prof. Madelung that the covariance does exist all the same. The given transformation is linear, but it is not of the usual vector analytical nature and thus his system does not come under the viewpoint discussed here.}

   In the present paper, the author discusses the problem from a somewhat different point of view. Without heuristically trying to find analogies with the classical field equations, he studies the transformation properties of the Dirac equation in general by using a formalism which seems to be well-adapted to the problem and is based upon the ``quaternions'' introduced by Hamilton. The quaternion calculus has never really been adopted in physics. Uniting vector and scalar multiplication into one operation proved to be insufficiently flexible for the purposes of vector calculation. And later, facing the magnificent unifying and generalizing viewpoints of tensor analysis, the operation with quantities based on special properties of three- or four-dimensional space had to be withdrawn in favor of a purely component-based representation. Nevertheless, there remains the fact that the quaternions can offer considerable practical ease in treating the general transformations of the Lorentz group, partly because they enable any given Lorentz transformation to be described fairly simply, and partly because those constructions, which in terms of tensor calculus are to be interpreted as invariant or covariant quantities, are also characterized by special and easily describable properties in quaternion calculus.

   The author arrived at the present research through the fact that in his doctoral dissertation written 10 years ago\footnote{``The Function Theoretical Relationships of the Maxwell Aether Equations,'' Publishing House of Josef Nemeth, Budapest 1919. (Due to the difficulties of the post-war period, the essay was published in 50 lithographed copies only.) Inaugural dissertation at the University of Szeged, Hungary. \emph{Editorial note:} This dissertation is now available in paper \cite{LAN1919-} and electronic \cite{HURNI2004-} forms, together with two commentaries \cite{GSPON1998A, GSPON2005-}.} he had dealt with
the demonstration of the connection between quaternions and Lorentz transformations, and he had pointed out how these can be used to formulate the laws of the special relativity theory formally in a very simple way, especially when applied to the electromagnetic field. Later, he no longer followed this line; however, an accidental observation led him to extend these investigations. Initially they were performed with respect to Maxwell's equations --- later to the Dirac equation by using the same mathematical tools, and indeed this course has proven to be correct and natural. The observation was that the formerly discussed equation system, which had been considered as a generalization of Maxwell's equations, is, interestingly, completely equivalent to Dirac's equation for the case where the mass term disappears. The inclusion of this term does not present any problem. Not only does it permit a fairly simple overview of the transformation properties of the functions, but it also offers new prospects concerning the tensor analytic meaning of the Dirac equation and may possibly lead to completely new viewpoints. Indeed, the following developments are of such almost trivial simplicity, and the direction of the progression is traced out so clearly, that one cannot help the feeling of having found a ``via regia'' to penetrate into the essence of Dirac's equation in a way that is more suitable than that allowed by the general operator method which is possibly unsuitable for that problem.

   Before beginning to discuss our actual problem, we will briefly summarize the foundations of the less well-known quaternion calculus and give an outline of the most important methods and results of the above-mentioned doctoral dissertation (Sections 2, 3, 4).

\section{The quaternions}

By a quaternion $Q$ we understand the combination of four quantities, ``components,'' in the form:
\begin{equation*}\label{1}
Q = X j_x + Y j_y + Z j_z + T j_l .
\tag{1}
\end{equation*}
The quantities $(j_x, j_y, j_z, j_l)$ are four ``unit vectors.'' For the fourth component\footnote{In the pre-relativistic period of Hamilton the fourth quaternion unit had, of course, no relationship to physical time and was added as a common scalar to the three spatial units. With regard to the applications we are interested in, we have immediately introduced the fourth dimension as a time dimension.} we chose the notation $l$ instead of $t$ because we want to reserve $t$ for the real time, whereas:
\begin{equation*}\label{2}
l = ict .
\tag{2}
\end{equation*}

   Along with the self-understood operation of addition, multiplication is the fundamental operation. This is defined by the multiplication of the unit vectors. We fix the relations:
\begin{equation*}\label{3}
\left.
\begin{array}{c}
j_x j_y = j_z , \quad j_y j_x = - j_x j_y , \quad j_x j_l = j_l j_x = j_x , \\
j_x^2 = j_y^2 = j_z^2 = - j_l^2 = -j_l
\end{array}
\right\}
\tag{3}
\end{equation*}
The remaining equations are obtained by cyclically interchanging $x, y, z$. The fourth unit vector $j_l$, in the direction of the imaginary ``time axis,'' behaves like the ordinary unit. Thus one can also write $j_l = 1$; in other words, $j_l$  can be ignored as a factor.

   The multiplication is associative but not commutative. Instead of the simple commutative law, here we have the law:
\begin{equation*}\label{4}
G F = \CON{ \CON{F} \, \CON{G} }
\tag{4}
\end{equation*}
or
\begin{equation*}\label{4a}
\CON{G F} = \CON{F} \, \CON{G} ,
\tag{4a}
\end{equation*}
where the bar means the following: one goes over to the ``conjugate'' of the quaternion, that is to say one gives the spatial components --- the space part as we shall call them --- opposite sign:
\begin{equation*}\label{5}
\CON{F} = -X j_x - Y j_y - Z j_z + T .
\tag{5}
\end{equation*}
For an arbitrary number of factors, the conjugate of the product is obtained by writing the sequence of factors in reverse order and taking the conjugate of each factor.

   It is easy to see that the quantity $F\CON{F}$ is simply a number (the spatial components~=~0).

\section{Four-dimensional rotations and quaternions}

Multiplication of quaternion $F$ by quaternion $p$ in the sense of
\begin{equation*}\label{6}
F' = p F
\tag{6}
\end{equation*}
can be conceived as a transformation of the line segment $F$ if the quaternion is represented as a vector in the four-dimensional space. The matrix of this transformation reads:
\begin{gather*}\label{7}
\begin{pmatrix}
p_4 & -p_3 & p_2 & p_1 \\
p_3 & p_4 & -p_1 & p_2 \\
-p_2 & p_1 & p_4 & p_3 \\
-p_1 & -p_2 & -p_3 & p_4 
\end{pmatrix}  .
\tag{7}
\end{gather*}
(Here the components of the quaternion $p$ are denoted by numbers instead of letters.) If we require that the norm of $p$ have the form
\begin{equation*}\label{8}
p \CON{p} = p_1^2 + p_2^2 + p_3^2 + p_4^2 = 1 ,
\tag{8}
\end{equation*}
then it is seen that the transformation is an orthogonal one. We shall denote an orthogonal transformation of this kind briefly as a ``$p$-transformation.'' From the associative law of multiplication, it follows that the $p$-transformations form a ``group'' which is contained in the general orthogonal transformations.

   A second group will be obtained by post-multiplying with the quaternion (instead of pre-multiplying):
\begin{equation*}\label{9}
F' = F q .
\tag{9}
\end{equation*}
This ``$q$-transformation''\footnote{In view of the completely different subject matter, this notation should not be confused with the customary one used for the generalized coordinates and momenta.} has the following matrix:
\begin{gather*}\label{10}
\begin{pmatrix}
q_4 & q_3 & -q_2 & q_1 \\
-q_3 & q_4 & q_1 & q_2 \\
q_2 & -q_1 & q_4 & q_3 \\
-q_1 & -q_2 & -q_3 & q_4 
\end{pmatrix}  .
\tag{10}
\end{gather*}
The two groups mutually complete each other by supplying in their composition the most general group of orthogonal transformations. Thus an arbitrary orthogonal transformation can be written in the form
\begin{equation*}\label{11}
F' = p F q ,
\tag{11}
\end{equation*}
given that:
\begin{gather*}\label{12}
p \CON{p} = 1 , \quad q \CON{q} = 1.
\tag{12}
\end{gather*}

   An arbitrary four-dimensional rotation is characterized by six parameters. In fact we have six quantities in the two quaternions since their lengths are normalized to 1.

   A subsystem of the general transformations is formed by the purely spatial rotations which do not change the time axis. Then obviously the same relationship holds between $\CON{F}' $ and $\CON{F}$ as holds between $ F ' $ and $F$. Since a reversal of the factors in \eqref{11} yields:
\begin{equation*}\label{11a}
\CON{F'} = \CON{q} \CON{F} \CON{p} ,
\tag{11a}
\end{equation*}
we must have:
\begin{gather*}\label{13}
p = \CON{q} , \quad q = \CON{p}.
\tag{13}
\end{gather*}
Accordingly, purely spatial rotations are characterized by the quaternion at the back being equal to the conjugate of the quaternion at the front.

   In the four-dimensional space of reality, one of the dimensions is imaginary. Accordingly, an orthogonal transformation cannot contain real coefficients exclusively, and we should consider both characteristic quaternions to be complex quantities. On the other hand, they cannot be arbitrary complex quantities. Namely, only real Lorentz transformations can be allowed because only these transform real $(x, y, z, t)$ into real $(x', y', z', t')$ again. This constraint can be characterized as follows: Let us consider the ``position vector:'' $R=(x, y, z, l)$. Let us change over to the conjugate complex quantity which we will denote by an ``asterisk'' ($^*$). As $l$ is imaginary, only the time component of $R$ will change its sign. Hence we have
\begin{equation*}\label{14}
R^* = - \CON{R} .
\tag{14}
\end{equation*}
This peculiarity must remain valid in the primed system as well. Hence, if
\begin{equation*}\label{15}
R' = p R q ,
\tag{15}
\end{equation*}
then:
\begin{equation*}\label{16}
p^* R^* q^* = - \CON{q} \CON{R} \CON{p}
\tag{16}
\end{equation*}
and so:
\begin{gather*}\label{17}
p^* = \CON{q} , \quad q^* = \CON{p} .
\tag{17}
\end{gather*}
Accordingly, real Lorentz transformations are characterized by the quaternion at the back being equal to the complex conjugate and ``overbarred'' value of the quaternion at the front. This means that it is enough to give $p$, and this value already determines $q$. So a real Lorentz transformation can be written in the form
\begin{equation*}\label{18}
F' = p F \CON{p}^* .
\tag{18}
\end{equation*}
In the case of spatial rotations the quaternion at the back must be equal to $\CON{p} $ and therefore it is seen that spatial rotations always belong to real $p$.

\section{The Hamiltonian operator}

Let us introduce the following differential operator (``gradient''):
\begin{equation*}\label{19}
\nabla = \left( j_x \frac{\partial ~}{\partial x} +j_y \frac{\partial ~}{\partial y} +j_z \frac{\partial ~}{\partial z} + \frac{\partial ~}{\partial l} \right) ,
\tag{19}
\end{equation*}
which we will denote as ``Hamiltonian operator.''\footnote{In Hamilton's works we can find only the spatial part of this operator. Strictly taken we should, therefore, speak of an ``extended Hamiltonian operator.'' For the sake of brevity, however, we forgo this, especially since this extension is quite obvious for the relativistic application of quaternions.} Under coordinate transformations it behaves like a vector.

   We shall apply this operation to a quaternion $F$ in the following form:
\begin{equation*}\label{20}
\CON{\nabla} F = \left( - \frac{\partial ~}{\partial x} j_x - \frac{\partial ~}{\partial y} j_y - \frac{\partial ~}{\partial z} j_z + \frac{\partial ~}{\partial l} \right) ( X j_x + Y j_y + Z j_z + T ) .
\tag{20}
\end{equation*}
Let us put:
\begin{equation*}\label{21}
\CON{\nabla} F = 0 ,
\tag{21}
\end{equation*}
so we carry out to some extent the same process in four dimensions, as in the complex function theory in two dimensions, if we write down the Cauchy-Riemann differential equations by putting:
\begin{equation*}
\left( \frac{\partial ~}{\partial x} + i \frac{\partial ~}{\partial y} \right) (u+iv)= \frac{\partial u}{\partial x} - \frac{\partial v}{\partial y} + i \left( \frac{\partial u}{\partial y} + \frac{\partial v}{\partial x} \right) = 0 .
\notag
\end{equation*}
It was this formal analogy which motivated me at the time to investigate equation \eqref{21}.

   If we decompose the above into components, we have the following system:
\begin{equation*}\label{21a}
\left.
\begin{aligned}
 \dfrac{\partial X}{\partial l} - \dfrac{\partial T}{\partial x} +
 \dfrac{\partial Y}{\partial z} - \dfrac{\partial Z}{\partial y} =& 0 , \\
 \dfrac{\partial Y}{\partial l} - \dfrac{\partial T}{\partial y} +
 \dfrac{\partial Z}{\partial x} - \dfrac{\partial X}{\partial z} =& 0 , \\ 
 \dfrac{\partial Z}{\partial l} - \dfrac{\partial T}{\partial z} +
 \dfrac{\partial X}{\partial y} - \dfrac{\partial Y}{\partial x} =& 0 , \\ 
 \dfrac{\partial X}{\partial x} + \dfrac{\partial Y}{\partial y} +
 \dfrac{\partial Z}{\partial z} + \dfrac{\partial T}{\partial l} =& 0 .
 \end{aligned}
\quad \right\} 
\tag{21a}
\end{equation*}

   These equations are closely connected with Maxwell's equations for empty space. There we have:
\begin{equation*}\label{22}
\left.
\begin{aligned}
\frac{1}{c} \frac{\partial \mathcal{E}}{\partial t}
           - \text{rot} \, \mathcal{H} &= 0 ,       \\
\frac{1}{c} \frac{\partial \mathcal{H}}{\partial t}
           - \text{rot} \, \mathcal{E} &= 0 ,       \\
             \text{div} \, \mathcal{E} &= 0 ,       \\
             \text{div} \, \mathcal{H} &= 0 . 
\end{aligned}
\right\}
\tag{22}
\end{equation*}
Subtracting the second equation multiplied by $i$ from the first, we obtain the complex equation:
\begin{equation*}\label{23}
\frac{1}{c} \frac{\partial (\mathcal{H}+i\mathcal{E})}{ i \partial t} - \text{rot} (\mathcal{H}+i\mathcal{E}) = 0 ,
\tag{23}
\end{equation*}
in which only the combination:
\begin{equation*}\label{24}
\mathcal{F} = \mathcal{H} + i \mathcal{E}
\tag{24}
\end{equation*}
occurs. In the same way, we can combine the last two equations in the form:
\begin{equation*}\label{25}
\text{div} ( \mathcal{H} + i \mathcal{E} )=0 .
\tag{25}
\end{equation*}

   If we denote the components of $\mathcal{F}$ by $X, Y, Z$ (which now also should be considered as complex numbers) and write down the equations component by component, it becomes clear that the system obtained is identical with equation \eqref{21a} if we put $T = 0$ in the latter.

   Therefore equation \eqref{21}, which is also Lorentz invariant and also supplies the wave equation for all components --- just like Maxwell's equations --- can be conceived as a natural formal extension of the latter. The wave equation is obtained by a second application of the $\nabla$-operation in the following form:
\begin{equation*}\label{26}
\nabla ( \CON{\nabla} F) = (\nabla \CON{\nabla}) F = 0.
\tag{26}
\end{equation*}
Namely, the $\nabla\CON{\nabla} $ operator is evidently a scalar and identical with the Laplacian $\Delta$ in four dimensions:
\begin{equation*}\label{27}
\nabla \CON{\nabla} = \Delta = \frac{\partial^2}{\partial x^2} + \frac{\partial^2}{\partial y^2} + \frac{\partial^2}{\partial z^2} +\frac{\partial^2}{\partial l^2} .
\tag{27}
\end{equation*}

\section{Dirac's equation for the case of vanishing mass}

If in \eqref{21a} we add the second equation multiplied by $i$ to the first, likewise the fourth to the third equation, then we have two equations which contain only the combinations $X+iY$ and $Z + iT$:
\begin{equation*}\label{28}
\left.
\begin{aligned}
  \frac{\partial (X+iY)}{\partial l} +i \frac{\partial (Z+iT)}{\partial x}
- \frac{\partial (Z+iT)}{\partial y} -i \frac{\partial (X+iY)}{\partial z} & = 0 , \\
  \frac{\partial (Z+iT)}{\partial l} +i \frac{\partial (X+iY)}{\partial x}
+ \frac{\partial (X+iY)}{\partial y} +i \frac{\partial (Z+iT)}{\partial z} & = 0 .
\end{aligned}
\right\}
\tag{28}
\end{equation*}
Of course, these two equations cannot replace the original four. The eight Maxwell equations were halved to four by changing over from real to complex quantities. Here, however, $X, Y, Z, T$ themselves are already complex. However, we can write two further equations by taking into account the system of equations for the conjugate complex quantities and performing the same operations there as well. Then obviously we should mark all components with an asterisk and also take into account that the term with $\partial / \partial l$ changes its sign. Thus for the quantities marked with an asterisk, we have the same equations but with an opposite sign in the first term.

   Let us introduce the following assignments:
\begin{equation*}\label{29}
\left.
\begin{aligned}
X+iY & = \psi_4 , & \quad X^* + i Y^* & = \psi_2 , \\
Z+iT & = \psi_3 , & \quad Z^* + i T^* & = \psi_1 ,
\end{aligned}
\right\}
\tag{29}
\end{equation*}
then it is seen immediately that we have Dirac's equations in the same form as they are explicitly written in Weyl's textbook\footnote{Hermann Weyl, Group Theory and Quantum Mechanics (Publishing House of S.\ Hirzel, Leipzig, 1928) p.\ 171.} if we neglect the mass term. From this we can see that the $\nabla $ operator can substitute fully for the Dirac operator, and this has the advantage that the Hamiltonian operator is in closer connection with vector analytical quantities.

   We note that the association of Dirac's equations to those of Maxwell has been performed by C.\ G.\ Darwin\footnote{loc.\ cit.; see footnote \ref{Darwin}.} in a similar way except that we have to put $T = 0$ in the Maxwellian case. Then the problem arises that $\psi_1$ and $\psi_3$ are not independent of each other; rather, the one is determined by the other. Equation \eqref{21}, on the other hand, allows exactly as many degrees of freedom as the Dirac equations, and they are in fact equivalent to them.

   Before starting to discuss the full Dirac equations, we should like to point out an important characteristic of equation \eqref{21} to which we will return later. The equations obviously do not unequivocally prescribe the transformation of quantity $F$ under a Lorentz transformation. In fact, without any rotation of the axes, from the associative law of multiplication it already follows that a transformation of the form:
\begin{equation*}\label{30}
F ' = F k
\tag{30}
\end{equation*}
(with k an arbitrary quaternion) is possible without any change of the system. In the case of a Lorentz transformation:
\begin{equation*}\label{31}
\nabla ' = p \nabla \CON{p}^*
\tag{31}
\end{equation*}
generally one can only say of the transformation of $F$ that:
\begin{equation*}\label{32}
F ' = p F k
\tag{32}
\end{equation*}
must hold where $k$ may still be an arbitrary quaternion.

   In the special case of Maxwell's equations, this ambiguity does not appear because from the very beginning $T = 0$ and it has to be demanded that $T ' = 0$ holds in the new system as well. Then the rotation imposed on $F$ by transformation \eqref{32} must also be a purely spatial one and we have:
\begin{equation*}\label{33}
k = \CON{p} .
\tag{33}
\end{equation*}

   We know that the electromagnetic field strength transforms in the four-dimensional formalism like an antisymmetric tensor (a six-vector). From this we can draw the following conclusion: If we know that a complex quaternion is transformed as:
\begin{equation*}\label{34}
p F \CON{p} ,
\tag{34}
\end{equation*}
then the time component is an invariant of the transformation while the real and imaginary parts of the spatial components can be considered as forming a six-vector. Of course, we could also give a direct proof for this type of transformation of an antisymmetric tensor, but here we shall dispense with this for the sake of brevity.

\section{Dirac's equation in the case of imaginary mass}

It is natural to try to introduce the mass term by introducing a term proportional to $F$ instead of setting the right-hand side of equation \eqref{21} equal to 0. However, it should be taken into consideration that it is not the repetition of the $\CON{\nabla}$-operation but the operation $\nabla\CON{\nabla}$ which gives the wave equation. On the other hand, we can make use of the specific characteristic of the four-dimensional space-time continuum, that:
\begin{equation*}\label{35}
\CON{\nabla} = - \nabla^* .
\tag{35}
\end{equation*}

   We now make the assumption:
\begin{equation*}\label{36}
\CON{\nabla} F = \alpha i F^* ,
\tag{36}
\end{equation*}
where $\alpha$ is any real number. With regard to rule \eqref{35} it follows from this equation that:
\begin{equation*}\label{37}
\nabla F^* = \alpha i F ,
\tag{37}
\end{equation*}
and so, applying the operation $\nabla $ to the first equation, we have:
\begin{equation*}\label{38}
\nabla \CON{\nabla} F = - \alpha^2 F .
\tag{38}
\end{equation*}
Thus we really obtain the required Schr\"odinger equation, but with the opposite sign of the mass term, which would only be possible in the case of imaginary mass.  Later we shall see that this change of sign is well-founded and cannot be eliminated from the equation by introducing numerical factors. Although an imaginary mass obviously has no real meaning, it is of heuristic value to discuss equation \eqref{36} briefly.  Let us combine the equations in pairs again, exactly as in equations \eqref{28}, and let us introduce the quantities $\psi$ in the sense of the assignment \eqref{29}. Then we obtain exactly the Dirac equation if we take the imaginary value:
\begin{equation*}\label{39}
m = - \frac{h}{2 \pi c} \alpha i
\tag{39}
\end{equation*}
for the mass $m$.

   In the case of a Lorentz transformation, the transformation law of quaternion $F$ is easily found. Let us set:
\begin{equation*}\label{40}
F ' = x F y ,
\tag{40}
\end{equation*}
then we have:
\begin{equation*}\label{41}
\CON{\nabla}' F ' = p^* \CON{\nabla} \CON{p} \, x F y .
\tag{41}
\end{equation*}
Hence it follows that:
\begin{gather*}\label{42}
x^* = p^* , \quad x = p , \quad y = y^* 
\tag{42}
\end{gather*}
must hold if equation \eqref{36} is to remain valid in the transformed system as well. The second condition is equivalent to the first. The third condition states that $y$ must be a real quaternion.

   Accordingly, the law of the transformation of $F$ reads:
\begin{equation*}\label{43}
F ' = p F r ,
\tag{43}
\end{equation*}
with an arbitrary real $r$, whose length we assume to be normalized to 1. It is obvious that no invariant meaning can be assigned to a quantity which behaves in this manner under a transformation. In the case of purely spatial rotations $p$ is real; thus we could put $r =\CON{p}$ and $F$ would be a vector. In the case of a general Lorentz transformation, however, such an assignment is impossible.

   This is not remarkable. When writing down equation \eqref{36}, we make essential use of the characteristic of the space-time continuum that one of the dimensions is imaginary. In a purely real four-dimensional manifold, an equation of the type \eqref{36} would lose its meaning.  However, the concepts of tensor analysis are formed in such a way that they never utilize the reality of the quadratic fundamental form, and they consider the peculiar values $+3,-1$ for the index of inertia of actual space-time as a coincidence. Here, however, it is just these peculiarities that are essentially utilized.

   Even if $F$ itself has no invariant meaning, it can be used to construct quantities having such a meaning. First we have an invariant:
\begin{equation*}\label{44}
F ' \CON{F}' = p F r \CON{r} \CON{F} \CON{p} = p ( F \CON{F} ) \CON{p} = F \CON{F} .
\tag{44}
\end{equation*}
This appears as a complex quantity and is thus equivalent to two real invariants, just the two invariants of the Dirac theory. Furthermore the quantity:
\begin{equation*}\label{45}
F \CON{F}^*
\tag{45}
\end{equation*}
forms a vector. Indeed, the transformation law reads:
\begin{equation*}\label{46}
F' \CON{F}' {}^* = p F r \CON{r}^* \CON{F}^* \CON{p}^* = p ( F \CON{F}^* ) \CON{p}^* ,
\tag{46}
\end{equation*}
since $r$ is real. This vector, the spatial part of which is purely imaginary and the time part purely real, is identical with the ``current vector of probability'' of the Dirac theory.

   There exist no further covariant constructions unless a restriction is introduced for $r$, which cannot be made without arbitrariness.

\section{Dirac's equation in the case of real mass}

It is also possible to arrive at the Schr\"odinger equation in the following more general way. Instead of supposing that $\CON{\nabla} F$ is proportional to a quantity formed from $F$, we introduce a new quantity $G$.
\begin{equation*}\label{47}
\CON{\nabla} F = \alpha G .
\tag{47}
\end{equation*}
If we want:
\begin{equation*}\label{48}
\nabla G = \beta F ,
\tag{48}
\end{equation*}
then obviously we have:
\begin{equation*}\label{49}
\nabla \CON{\nabla} F = - \alpha \beta F .
\tag{49}
\end{equation*}
Here the product $\alpha \beta $ can be either positive or negative. By attaching a factor to $G$, the constant on the right-hand side can be made equal or the negative of each other, whereby we have reduced the system to two normal forms, namely either:
\begin{equation}\label{50}
\left.
\begin{aligned}
\CON{\nabla} F & = \alpha G , \\
\nabla G & = \alpha F , 
\end{aligned}
\right\}
\tag{50}
\end{equation}
or
\begin{equation}\label{51}
\left.
\begin{aligned}
\CON{\nabla} F & = \alpha G , \\
\nabla G & = - \alpha F .
\end{aligned}
\right\}
\tag{51}
\end{equation}

   Let us first consider the second case. If we change over to the conjugate complex quantity in the second equation then we can also replace \eqref{51} with:
\begin{equation}\label{51a}
\left.
\begin{aligned}
\CON{\nabla} F & = \alpha G , \\
\CON{\nabla} G^* & = \alpha F^* .
\end{aligned}
\right\}
\tag{51a}
\end{equation}
Now we can add these two equations and write:
\begin{equation*}\label{52}
\CON{\nabla} (F + G^* ) = \alpha (F + G^* )^*
\tag{52}
\end{equation*}
or
\begin{equation*}\label{53}
\CON{\nabla} (F + iG^* ) = i \alpha (F + i G^* )^* ,
\tag{53}
\end{equation*}
and in this way we have arrived at our former equation combination \eqref{36} for the combination $H=F+iG^*$.

   However, in the first case, corresponding just to a real mass, a unification of this type, which should lead to an equation for only one single quantity, is not automatically possible.\footnote{Later we shall see, however, that there exist corresponding combinations here as well if quaternions are introduced as factors.}  In the following we shall deal with this case.

It is useful to write our fundamental equation in the form:
\begin{equation}\label{54}
\left.
\begin{aligned}
\CON{\nabla} F & = \alpha G^* , \\
\CON{\nabla} G & = - \alpha F^* .
\end{aligned}
\right\}
\tag{54}
\end{equation}
Now we can obtain the Dirac equation again if we write the combination (I + i\,II, III + i\,IV) in both the first and second system. Then we have to make the assignment:
\begin{equation}\label{55}
\left.
\begin{aligned}
X_1 + iY_1 & = \psi_4 , & \quad X_2^* + iY_2^* & = \psi_2 , \\
Z_1 + iT_1 & = \psi_3 , & \quad Z_2^* + iT_2^* & = \psi_1 ,
\end{aligned}
\right\}
\tag{55}
\end{equation}
and we have for the mass:
\begin{equation*}\label{56}
m = - \frac{h}{2 \pi c} \alpha .
\tag{56}
\end{equation*}
The indices 1 and 2 refer to quaternions $F$ and $G$, respectively. However, the system \eqref{54}, consisting of eight equations, is still not exhausted. Now we can exchange the roles of $F$ and $G$ and make the assignment:
\begin{equation}\label{57}
\left.
\begin{aligned}
X_2 + iY_2 & = \sigma_4 , & \quad X_1^* + iY_1^* & = \sigma_2 , \\
Z_2 + iT_2 & = \sigma_3 , & \quad Z_1^* + iT_1^* & = \sigma_1 .
\end{aligned}
\right\}
\tag{57}
\end{equation}
A Dirac equation system holds also for the values of $\sigma$ which are independent of $\psi$.  This equation system is completely identical with the first one, except that the signs are opposite in the mass term due to the change of sign in the second equation \eqref{54}. These two systems altogether are now equivalent to system \eqref{54}.\footnote{A duplication of the Dirac equation can also be found with the Madelung model (loc.\ cit.) where eight complex equations are written down.}

   In the case of a Lorentz transformation \eqref{31}, the transformation of $F$ and $G$ is easily obtained in the following form. It reads:
\begin{equation}\label{58}
\left.
\begin{aligned}
F' & = p F k , \\
G' & = p G k^* ,
\end{aligned}
\right\}
\tag{58}
\end{equation}
where $k$ may be an arbitrary quaternion.

   Here, too, the transformation displays a peculiar uncertainty in $\psi$ for which no correspondence can be found in the Dirac theory because there the transformation of the quantities is unequivocal. This can be explained as follows: For a transformation of $(X_1, ... , T_1)$, $(X_2, ... , T_2)$, after conversion to $\psi$ and $\sigma$, the latter quantities will generally be mixed. If, however, we presuppose only one Dirac equation, e.g., that containing $\psi$, then only those transformations come into consideration in which the quantities $\sigma$ do not appear, i.e., those in which $\psi$ are transformed into each other. This means that here we distinguish a subsystem in equation system \eqref{54} and require that this subsystem should be transformed into itself. Thereby the transformation of the quantities $\psi$ will be unequivocally defined (up to a common factor), but the so-distinguished subsystem will not get an invariant meaning, whereas this can be expected of the whole system. In fact, this will be confirmed and we shall be able to find a general covariant formulation for the extended system. However, before coming to this, it is interesting to have a closer look at the Dirac subsystem and its specific transformation characteristics --- in view of the importance of Dirac's theory.

\section{Unified derivation of the covariants of Dirac's theory}

In accordance with the above, we now presuppose the equation system for $\psi$ alone and allow only those transformations which convert $\psi$ into themselves. We can easily show that, under this condition, the indefinite quaternion $k$ of equation \eqref{58} can practically be set equal to~1. We only have to consider that now $X+iY$ and $Z+iT$ may only be transformed in this combination; otherwise, the quantities $X-iY$ and $Z-iT$ which can be deduced from the quantities $\sigma$ would also be required. It is easy to show that this means the following for the transformation matrix: we divide the square of the matrix into four small squares with a horizontal and a vertical line across the center. Then in each of these squares the two terms along the diagonal must be equal and the other two the negative of each other. If we check the matrix \eqref{7} of a $p$-transformation for this characteristic, we shall see that this condition is actually satisfied. In case of a $q$-transformation with a matrix \eqref{10}, however, this is true for two squares only and it does not hold for the remaining two. Therefore, we have to set $k_1$ and $k_2 = 1$ for the post-multiplying $k$-transformation and there remains only a multiplication of all $\psi$; with the same complex number.

   This transformation is trivial because of the linearity of the equation. It is natural to eliminate this by putting $k = 1$. Then a similarity transformation of this kind, which is possible even without any rotation of the coordinate system, is excluded and the transformation of $F$ and $G$ becomes unequivocal. However, normalization does not go this far in quantum mechanics. There, only the absolute value of each complex factor is normalized but not its phase. This is based upon the fact that for quantum mechanics it is only the ``hermitian operations'' which play a role --- they are not influenced by this phase. Hence, if we put $k = 1$, then we consider a more restricted group of transformations than those allowed by quantum mechanics. In this way we surely obtain all covariants that have a quantum mechanical meaning and possibly even more. We can proceed as follows: We check all covariant constructions of the restricted transformation group $k = 1$ and eliminate afterwards those which are not compatible with the mentioned ``phase transformation.''

   Now let us consider unequivocally the following transformation:
\begin{equation}\label{59}
\left.
\begin{aligned}
F' & = p F , \\
G' & = p G . 
\end{aligned}
\right\}
\tag{59}
\end{equation}
We do not know the quantities $F$ and $G$, only those combinations of their components which occur in equations \eqref{55}. Instead of these quantities we introduce a single quaternion $H$ which we assign to the quantities $\psi$ in the following way:
\begin{equation}\label{60}
\left.
\begin{aligned}
X + iY & = \psi_4 , & \quad X^* + iY^* & = \psi_2 , \\
Z + iT & = \psi_3 , & \quad Z^* + iT^* & = \psi_1 .
\end{aligned}
\right\}
\tag{60}
\end{equation}
It is easy to calculate that $H$ is composed of $F$ and $G$ as follows
\begin{equation*}\label{61}
2H = F + G + i (F - G ) j_z .
\tag{61}
\end{equation*}
Let us put:
\begin{equation*}\label{62}
G + F = I , \quad G - F = K ,
\tag{62}
\end{equation*}
then
\begin{equation*}\label{61a}
2H = I - i K j_z .
\tag{61a}
\end{equation*}

   We can easily deduce a differential equation for $H$. Namely, if we apply the operation $\CON{\nabla}$ to \eqref{61a} --- keeping in mind that equations \eqref{54} hold between $I$ and $K$ just as for $F$ and $G$ --- then we have:
\begin{equation*}
\begin{aligned}
2 \CON{\nabla} H & = \alpha (K^* + i I^* j_z ) = \alpha (-K^* j_z + i I^* ) j_z  \\
& = \alpha i( I^* + i K^* j_z ) j_z = 2\alpha i H^* j_z ,
\notag
\end{aligned}
\end{equation*}
that is:
\begin{equation*}\label{63}
\CON{\nabla} H = \alpha i H^* j_z .
\tag{63}
\end{equation*}
This equation is --- as one can easily verify --- equivalent\footnote{Of course, either $j_x$ or $j_y$ could occur instead of $j_z$ on the right-hand side, and then we would only have to cyclically interchange $X,Y$ and $Z$ during the assignment of the components to $\psi$ in equations \eqref{60}.} to the Dirac
equation between the quantities $\psi$.\footnote{\emph{Editorial note:} Equation \eqref{63}, to be called the \emph{Dirac-Lanczos equation}, is an important result of this paper. In Dirac's formulation the 4-complex-component electron field is taken as a $4 \times 1$ column vector $\Psi$, and the linear operators are $4 \times 4$ complex matrices.

   In Lanczos's formulation the same 4-complex-component field is a biquaternion $H \in \mathbb{B} \cong M_2(\mathbb{C}) \cong \cl_{1,2} \cong \cl_{3,0}$. The linear operators are then linear biquaternions functions of biquaternions, which are isomorphic to the algebra of $4 \times 4$ complex matrices  $M_4(\mathbb{C})  \cong M_2(\mathbb{B}) \cong \cl_{4,1}$.

   In both formulations the operator space has $4 \times 4 \times 2 = 32$ dimensions over the reals. The difference is that in the Dirac formulation the field is an abstract 4-component column vector, while in the Lanczos formulation the field is directly related to the algebraic structure of spacetime because any biquaternion $H = s + \vec{v}$ is the direct sum of a scalar $s$  and a 3-component vector $\vec{v}$.

   Lanczos's formulation is therefore more suitable than Dirac's for studying and demonstrating the ``classical'' aspects of the electron field, and for making comparisons with the Maxwell and Proca fields which are usually expressed in terms of scalars and vectors.

   Finally, in terms of Clifford algebras, the Dirac field $\Psi$ is a degenerate 8-real-component element of the 32-dimensional Clifford algebra $\cl_{4,1}$ (i.e., an element of an ideal of that algebra) while the Lanczos field $H$ is any 8-real-component element of the 8-dimensional Clifford algebra $\cl_{1,2} \cong \mathbb{B}$, which is therefore the \emph{smallest} algebra in which Dirac's electron theory can be fully expressed.

  For more details see Refs.~\cite{GSPON1998B,GSPON1994-,GSPON2001-,GSPON2002-}, where it is shown that the complex conjugation operation appearing on the right-hand side of Eq.~\eqref{63} is characteristic of the Fermionic character of the Dirac field.
}
If we had not had in mind an invariant formulation of the Dirac equation, then we could have started out with this equation from the very beginning and it could have formed the basis of our investigation.

   We can also write down for $H$ the transformation:
\begin{equation*}\label{64}
H' = p H ,
\tag{64}
\end{equation*}
and it can immediately be seen that:
\begin{equation*}\label{65}
H' \CON{H}' = p H \CON{H} \CON{p} = H \CON{H}
\tag{65}
\end{equation*}
is an invariant. If we separate the real and imaginary parts:
\begin{equation*}\label{66}
H \CON{H} = A + Bi ,
\tag{66}
\end{equation*}
then we obtain two invariants --- these are just the two fundamental invariants of the Dirac theory.

   It is also easy to see that the quantity:
\begin{equation*}\label{67}
H' \CON{H}'{}^* = p (H \CON{H} ) \CON{p}^*
\tag{67}
\end{equation*}
forms a vector. This vector represents the ``probability current'' in the Dirac theory, and its zero divergence was proved by Dirac.\footnote{Proc.\ Roy.\ Soc.\ {\bf 118}, 251, 1928.}

   Remarkably, this vector is supplemented by three others which can be obtained as follows: Let $V$ be an arbitrary vector, and let us form the following product:
\begin{equation*}\label{68}
\CON{H} V H^* .
\tag{68}
\end{equation*}
We shall prove that this is an invariant. In fact we have:
\begin{equation*}\label{69}
\CON{H}' V' H'{}^* = \CON{H} \CON{p} p V \CON{p}^* p^* H^* = \CON{H} V H^* .
\tag{69}
\end{equation*}
Now let us write this product in the form:
\begin{equation*}\label{70}
\CON{H} V H^* = \sum_{\alpha = 1}^4 (\CON{H} j_{\alpha} H^*) V_{\alpha} ,
\tag{70}
\end{equation*}
where $V_{ \alpha}$ denotes the components of $V$. However, if $\sum B_{ \alpha} V_{ \alpha}$ is an invariant, then the $ B_{ \alpha}$ necessarily represents the components of a vector. Consequently we obtain a vector with the components:
\begin{equation*}\label{71}
B_{\alpha} = \CON{H} j_{\alpha} H^* .
\tag{71}
\end{equation*}
In actuality, not only one but four vectors are obtained in this way, since the invariant \eqref{68} is a quaternion itself and is thus equivalent to four invariants because each coefficient of $j_i$ remains unchanged in itself. Let us write down the components of the four quaternions each in a line, one under the other. In this way we obtain a quadratic array containing the components of a vector in each column.

   Now the scheme \eqref{71} can be realized in a very simple way. Namely, we can consider it as an orthogonal transformation of $j_{ \alpha}$. In fact, it is obviously a real Lorentz transformation (since the quaternion at the back of this expression is the conjugate complex and barred quantity of the one at the front). Accordingly, we form first a $p$-matrix from $\CON{H}$, a $q$-matrix from $H^*$ and multiply these two matrices. Applying the resultant matrix to $j_{ \alpha}$, we obtain an array in which only the columns and rows are to be interchanged. The components of the individual vectors are, however, each contained in columns. Thus the rows of the matrix obtained immediately provide the four vectors. The fourth is identical with the current vector obtained earlier.

   The transformation matrix obtained is an orthogonal one. Thus we obtain four vectors at each point which are perpendicular to each other. One of them is the current vector. The length of all these vectors is the same.\footnote{This geometrical result is methodologically very interesting, especially in view of the Einstein theory of ``Distant parallelism'' which is just based on such ``local n-frames.''}

The square of the length is the same for  all four:
\begin{equation*}
= (H \CON{H}) ( H^* \CON{H}^* ) = A^2 + B^2 .
\notag
\end{equation*}
It is also easy to calculate the divergence of these vectors (which must be an invariant). We should then form:\footnote{\emph{Editorial note:} In equations \eqref{72} and \eqref{73} the parentheses are meant to specify the range of action of the differential operators ${\partial}/{\partial x_{\alpha}}$ and $\nabla$.}
\begin{equation*}\label{72}
\begin{aligned}
\sum \frac{\partial B_{\alpha}}{\partial x_{\alpha}} & 
  =  \sum \left( \frac{\partial}{\partial x_{\alpha}} \CON{H} j_a H^* \right)
  = ( \CON{H} \nabla ) H^* + \CON{H} ( \nabla H^* )  \\
& = ( \CON{ \CON{\nabla} H } ) H^* - \CON{H} ( \CON{\nabla} H )^*
  = U - \CON{U}^* ,
\end{aligned}
\tag{72}
\end{equation*}
where we put:
\begin{equation*}\label{73}
U = \bigl( \CON{ \CON{\nabla} H } \bigr)  H^*   .
\tag{73}
\end{equation*}
Since the differential equation \eqref{63} holds for $H$, it follows that:
\begin{equation*}\label{74}
U = - \alpha i j_z \CON{H}^* H^* = - \alpha i (A -iB) j_z ,
\tag{74}
\end{equation*}
and we obtain:
\begin{equation*}\label{75}
\sum \frac{\partial B_{\alpha}}{\partial x_{\alpha}} = -2 \alpha B j_z .
\tag{75}
\end{equation*}
This means that only the vector belonging to the third row of the matrix has a divergence which differs from zero.

   The remaining three orthogonal vectors --- among them the current vector --- are divergence free.

   We can also write the four vectors obtained in the form:
\begin{equation*}\label{76}
B^{(i)} = H j_i \CON{H}^* ,
\tag{76}
\end{equation*}
where $j_i$ is one of the four quaternion units. Only the third and fourth vectors are not affected by the phase transformation. The fourth is the previously mentioned Dirac current vector. The third,
\begin{equation*}\label{77}
B^{(3)} = H j_z \CON{H}^* ,
\tag{77}
\end{equation*}
is the one whose divergence differs from zero.

   We can find the tensors of the Dirac theory in a similarly systematic manner. Let us form:
\begin{equation*}\label{78}
H' j_{\alpha} \CON{H'} = p H j_{\alpha} \CON{H} \CON{p} ,
\tag{78}
\end{equation*}
where $j_{ \alpha}$ should be one of the three spatial quaternion units. (The fourth unit supplies only the already known invariant.) We already know that a quantity which behaves like this under transformation can be considered in its spatial part as an antisymmetric tensor if we separate the real and imaginary  parts (just as we saw in tile case of the electromagnetic field strength). The time part gives an invariant. This time part, however, is zero for all three constructions and therefore gives nothing.

   The phase transformation is only compatible with the quantity formed with $j_z$:
\begin{equation*}\label{79}
H j_z \CON{H} ,
\tag{79}
\end{equation*}
which is thus the only quantum mechanically allowed one. This antisymmetric tensor was introduced by C.\ G.\ Darwin.\footnote{Proc.\ Roy.\ Soc.\ {\bf 120}, 621, 1928.}

   There is another possible way to obtain a tensor.  Using the vectors $U$ and $V$ we form the invariant:
\begin{equation*}\label{80}
\CON{H}' U' \CON{V}' H' = \CON{H} \CON{p} p U \CON{p}^* p^* \CON{V} \CON{p} p H = \CON{H} U \CON{V} H .
\tag{80}
\end{equation*}
Let us write this in the following form:
\begin{equation*}\label{81}
\sum_{\mu , \nu} ( \CON{H} j_{\mu} \CON{j}_{\nu} H ) U_{\mu} V_{\nu} ,
\tag{81}
\end{equation*}
then we have a bilinear form the coefficients of which must be the components of a tensor:
\begin{equation*}\label{82}
T_{ik} = \CON{H} j_i \CON{j}_k H .
\tag{82}
\end{equation*}
Here, too, we obtain four tensors simultaneously since the components appear as quaternions and a tensor component can be separated from each unit vector.

   We could just as well have started from the invariant:
\begin{equation*}\label{83}
\CON{H}^* \CON{U} V H^*
\tag{83}
\end{equation*}
and, correspondingly, we would have arrived at the tensor:
\begin{equation*}\label{84}
T'_{ik} = \CON{H}^* \CON{j}_i j_k H^* .
\tag{84}
\end{equation*}
With this we have listed all those covariant constructions of zero, first and second degree for the restricted group $k = 1$, which are built quadratically from the fundamental quantities.

   The covariants of the Dirac theory have already been discussed in the literature.\footnote{See especially J. von Neumann, Zeits.\ f.\ Phys.\ {\bf 48}, 868, 1928.} However, the uniform development outlined here may outdo the methodological persuasive power of other descriptions by its clarity and simplicity.

\section{Failure of the current vector with respect to strict covariance}

Once again surveying our train of thought, we can state the following: We started from a larger equation system and found that the transformation of the functions is not unequivocally determined by the equations. However, we required that a given subsystem of the system transforms into itself and this requirement enabled us to eliminate the uncertainty of the transformation.

   However, if we only wish to investigate the Dirac equation, we could just as well have taken equation \eqref{63} as a basis from the very beginning. (We have derived this equation from the larger system for a certain combination of $F$ and $G$.)  We could have performed the assignment \eqref{60} and could have shown that equation \eqref{63} is indeed equivalent to the Dirac system. If we now take this path and want to calculate the transformation of the function $H$ immediately from this system, then we shall experience a peculiar discrepancy with our last results, and this is probably not without significance. Let us write the transformation of $H$ in the form:
\begin{equation*}\label{85}
H' = p H k ,
\tag{85}
\end{equation*}
then we obtain the following condition for quaternion $k$:
\begin{equation*}\label{86}
j_z k = k^* j_z .
\tag{86}
\end{equation*}
This condition by no means represents a strict limitation on $k$ that would allow, for instance, only a trivial similarity transformation. Rather, condition \eqref{86} formulates the requirement that the $x$ and $y$ parts of $k$ must be purely imaginary and its $z$ and $l$ parts purely real. Now if we also perform a natural length normalization, which excludes just the trivial similarity transformation, then a very important $3$-parameter group of transformations will still remain which certainly has nothing to do with the ``phase uncertainty'' of the $\psi$. Rather, one can observe that in the case of this transformation which is possible even without any rotation of the axes, the new $\psi$ can be expressed not only in terms of the previous $\psi$ but also in terms of their conjugates $\psi^*$.  As $\psi$ only plays the role of auxiliary quantities, no objective reasons can be found for excluding these transformations, even more so because these are absolutely normal transformations for the $(X, Y, Z, T)$ --- without the appearance of the conjugates. No reason can be found why we should not introduce these quantities instead of $\psi$ as fundamental quantities, though obviously the Hamiltonian operator is more closely connected with tensor analytical quantities than is the Dirac operator. If, however, we allow these transformations, then a large part of the covariants set up will be lost. Only the invariant $H\CON{H} $ and a vector will remain. However, this is not the current vector with zero divergence but vector $B^{(3)}$ whose divergence proved to be different from zero. In fact, we obtain:
\begin{equation*}\label{87}
H' j_z \CON{H}^* = p H k j_z \CON{k}^* \CON{H}^* \CON{p}^* = p H j_z \CON{H}^* \CON{p}^* ,
\tag{87}
\end{equation*}
because from \eqref{86} it follows that:
\begin{equation*}\label{88}
j_z = k^* j_z \CON{k} = k j_z \CON{k}^* .
\tag{88}
\end{equation*}
For the current vector, we cannot prove the invariance under this $k$-transformation,  Thus $B^{(3)}$ is the only vector which really seems to be strictly covariant. In the case of the other constructions (especially with the current vector), the covariance is the result of what is, after all, an arbitrary constraint.\footnote{Remark during revision: The author had not suspected that the transformation properties could change due to the introduction of the vector potential. Actually, this is exactly the case; thus the special group of transformations found here is lost with the extension of the system by the external field. The objection that was made against the customary transformation theory of the $\psi$ functions is thus invalid. For more details, see the paper ``On the covariant formulation of Dirac's equation'' to be published in the near future. (\emph{Editorial note:} See Ref.~\cite{LAN1929C}.)}

   This circumstance seems to suggest that the Dirac equation should be considered as a component of a larger system, instead of a system closed in itself.

\section{Covariant formulation of the doubled Dirac equation}

Let us now consider system \eqref{54} as a whole and let us study its transformation properties. As we have seen, the system is equivalent to two simultaneous Dirac equations which contain two independent groups of quantities: $\psi$ and $\sigma$. However, from the standpoint of the whole system, such a decomposition would be rather constrained and unnatural and of no advantage even as a mathematical aid, since it would amount to separating out a subsystem which is not internally preferred. We want to refrain from doing this and to be guided rather by the point of view that a covariant meaning should be attributed to the quantities $F$ and $G$ contained in the equations. This principle will lead us to remove the uncertainty of the transformation which lies in the arbitrariness of the quaternion $k$, whereas until now we attained this just by separating out a subsystem of equations which is not covariant in itself.

   Let us once again write down the transformation equations of our functions $F$ and $G$:
\begin{equation}\label{58'}
\left.
\begin{aligned}
F' & = p F k , \\
G' & = p G k^* ,
\end{aligned}
\right\}
\tag{58}
\end{equation}
where $k$ may be arbitrary and is only subject to the always possible, but natural, normalization $k\CON{k} = 1$. Before we come to the determination of $k$ on the basis of the covariance principle, let us note that certain invariants and covariants are possible even with $k$ left arbitrary. 

   Namely, first the two invariants:
\begin{equation*}\label{89}
F \CON{F} , \quad G \CON{G} ,
\tag{89}
\end{equation*}
which are equivalent to four real invariants.

   Then the two vectors can be formed:
\begin{equation*}\label{90}
F \CON{G}^* , \quad G \CON{F}^* ,
\tag{90}
\end{equation*}
so that, for example,
\begin{equation*}\label{91}
F' \CON{G}'{}^* = p F k \CON{k} \, \CON{G}^* \CON{p}^* = p (F \CON{G}^*) \CON{p}^* .
\tag{91}
\end{equation*}
The two vectors of \eqref{90} are, however, not independent of each other; rather, the second is equal to the conjugate complex and overbarred quantity of the first. Consequently, it is enough to consider $F\CON{G}^*$ alone.

   We might guess that this vector --- i.e., its real and imaginary
parts --- should correspond to the current vector, as it represents the analog construction of $H \CON{H}^*$. However, if we calculate its divergence using a method quite similar to that applied in \eqref{72}, we shall find this is not confirmed. Instead we shall have:
\begin{equation*}\label{92}
\operatorname{div} (F\CON{G}^*)
       = \Scal[\CON\nabla F \CON{G}^*]
       = \alpha [(F \CON{F}) + (G \CON{G})^*] .
\tag{92}
\end{equation*}

   This finding could be used as a negative factor against the interpretation suggested here. In fact, this objection would not be justified. This is because this vector, distinguished by the fact that it remains covariant even under an unlimited transformation of the equations, does not correspond at all to the current vector of the Dirac equation but to vector \eqref{77}, the divergence of which does not vanish even there. However, the Dirac current vector cannot unequivocally be made to correspond to a covariant formulation, as it is not characterized by invariant properties, and its covariance is due rather to an unnecessary limitation of the transformations --- as was found in the last section.

   In our formal introduction, we were able to find a tensor analytical interpretation for two types of quantities. If $ k =\CON{p}^*$, then we have to deal with a vector. If $ k =\CON{p} $, then we can consider the construction as a non symmetric tensor --- with its time component an invariant. Strangely enough, whichever choice we make both constructions occur. The difference is only that $F$ and $G$ exchange their roles.
Now we want to choose:
\begin{equation*}\label{93}
k = \CON{p} .
\tag{93}
\end{equation*}
By this choice we put $F$ as an antisymmetric tensor and $G$ as a vector. Now only covariant quantities occur in our equation. Also, both systems of equations are already covariant in themselves, not just the whole system.

   Let us examine now the first system. This system seems to be closely related to Maxwell's equations of the electromagnetic field. Here we are not thinking of the vacuum equations but of the complete equations with the current vector on their right-hand side. Namely, Maxwell's equation containing a current vector can be brought to the following form by the complex combination \eqref{24}:
\begin{equation*}\label{94}
\CON{\nabla} F = S^* ,
\tag{94}
\end{equation*}
where $S$ means the current vector considered as a quaternion. This is, however, just our first equation which is thus given a very simple interpretation with the aid of a classical analogy. Now we should only put $F_4 = 0$ for Maxwell's scheme but the occurrence of this term means only that the gradient of a scalar appears on the left-hand side. In addition, the complex character of our vector $G$ should be interpreted so that not only an electric but also a magnetic current must be taken into consideration.

   The emergence of the second equation is essentially new and unknown to the classical theory. This equation means that there is a ``feedback'' between current and field intensity, a reaction of the field to the current. The appearance of this equation should be looked upon as the actual effect that led to the discovery of the Schr\"odinger equation.

   The second equation also has a simple covariant meaning. An equation of this type is likewise well-known to us from electromagnetic field theory. If, namely, we write down the electromagnetic field by means of a vector potential  $\Phi_i$, regarding the $\Phi_i$ vector as a quaternion, then we obtain the equation:
\begin{equation*}\label{95}
\CON{\nabla} \Phi = - F^* ,
\tag{95}
\end{equation*}
which is just our second equation where we only have to consider that in our case $\Phi$ is to be treated as a complex quaternion. This implies that in addition to the customary ``rotation construction,'' the ``dual'' construction appears as well, in accordance with the occurrence of a ``magnetic vector potential.''

   Thus we can convert our equation system into the common language of physics either with the customary vectorial symbols or by using the terminology of tensor analysis. For this we only have to apply the corresponding equations of the electromagnetic field and substitute the corresponding quantities. The only difference is that the continuity condition for the current cannot be inferred from the general structure of the equations because of the occurrence of the redundant invariant which is not contained in the theory of Maxwell. This invariant is responsible for the fact that the disappearance of the divergence for the current vector is not a necessary consequence of the equations and that the equation system does not contain any internal correlations. (Absence of identities.)

   For the sake of clarity, we shall write down the resulting equations explicitly, in tensor analytical form. Then the following quantities appear:

   An antisymmetric tensor:  $F_{ik} = -F_{ki}$ (``electromagnetic field intensity''). Let the ``dual'' tensor be denoted by $\DUA{F}_{ik}$.\footnote{The dual assignment should be performed according to the following scheme:
\begin{equation*}\notag
\DUA{F}_{12} = i F_{34} , ... , \DUA{F}_{14} = i F_{23} , ...
\end{equation*}
(The dots indicate cyclic interchanging of 1, 2, 3.)
}

   Two vectors: $S_i$ and $M_i$ (``electric and magnetic current'').

   Two invariants: $S$ and $M$.

   We obtain altogether 16 equations which can be divided into two groups (A and~B). Group (A) contains two vector equations; group (B) contains an antisymmetric tensor equation (six equations) and two scalar equations.

   With the customary symbolism of tensor calculus, the equations read as follows:
\begin{equation*}\label{96}
\left.
\begin{array}{c}
(A) \quad \left\{
\begin{array}{c}
\dfrac{\partial S}{\partial x_i} + \dfrac{\partial F_{i \mu}}{\partial x_{\mu}}= \alpha S_i , \\ ~\\
\dfrac{\partial M}{\partial x_i} - \dfrac{\partial \DUA{F}_{i \mu}}{\partial x_{\mu}}= \alpha M_i ;
\end{array}
\right. \\
\\
(B) \quad \left\{
\begin{array}{c}
   \dfrac{\partial S_i}{\partial x_k}
    - \dfrac{\partial S_k}{\partial x_i}
    + \DUA{\left( \dfrac{\partial M_i}{\partial x_k}
    - \dfrac{\partial M_k}{\partial x_i} \right)}
    = \alpha F_{ik} , \\ ~~\\
\dfrac{\partial S_{\mu}}{\partial x_{\mu}} = \alpha S , \\ ~~\\
\dfrac{\partial M_{\mu}}{\partial x_{\mu}} = \alpha M .
\end{array}
\right.
\end{array}
\right\}
\tag{96}
\end{equation*}
Here the constant $\alpha$ is defined as
\begin{equation*}\label{97}
\alpha = \frac{2 \pi m c}{h} .
\tag{97}
\end{equation*}

   The whole system, just like the Dirac equations, displays a strong symmetry. However, it is questionable whether all quantities have a real meaning. So we might guess that the two scalars $S$ and $M$ do not appear in reality.  The first group of equations would then be fully equivalent to Maxwell's equations. The continuity condition for the current would be a consequence of the equations.  Despite this specialization, the previously mentioned difficulty experienced with the assignment made by Darwin\footnote{loc.\ cit.; see footnote \ref{Darwin}.} does not appear here. Namely, the two groups of  $\psi$-quantities ($\psi$ and $\sigma$) are assigned not to the field intensity alone but to the combination of field intensity and current vector simultaneously, and both types of quantities appear in each group [as can be seen from equations \eqref{55} and \eqref{57}]. This is why now there is no algebraic relationship between the quantities $\psi$ despite the specialization.

   Moreover, a further heuristic aspect can be derived from Maxwell's equations, namely, for the ``magnetic current vector'' $M_i$ which vanishes there. We must, however, take into consideration that here we can by no means dispose of quantities so freely as in common field theory.  Namely, due to the double coupling, there are no dependent and independent quantities here and the number of equations is just as large as the number of unknown quantities. It is just the characteristic feature of quantum mechanics that it operates with homogeneous equations (``eigenfunctions''), and the extraneous functions (energy, vector potential) appear not as ``right-hand sides'' of equations but as factors. If we put in our system certain quantities equal to zero, this would lead to overdetermination: the number of equations would be larger than that of unknown quantities. Then, accordingly, such a zero assumption should be compensated for by the omission of equations to avoid possible contradictions --- unless there are a number of identities, whereby the redundant equations would seem to be pure consequences of the others.

   From a mathematics and aesthetics viewpoint, this redundancy would be of considerable benefit. It cannot be denied that the strong symmetry of Maxwell's equations with respect to duality between electric and magnetic field intensities exists only as long as the equations are written down with the symbols of three-dimensional vector analysis. With respect to the four-dimensional tensor analytical description, there is an internal difference between the two systems of Maxwell's equations.  In reality, this difference manifests itself in the absence of magnetic current. In one of the systems, a normal divergence appears, whereas the other system contains the divergence of the ``dual'' field intensity. The tensor analytical formulation is faultless, yet it makes use of an accident in the four-dimensional space: the fact that in the case of a second-rank tensor, the ``dual'' tensor will be exactly of second-rank again. In principle, the emergence of this circumstance is logically not really satisfying.

   Now, if we put $M_i = 0$ for the magnetic current in our equations \eqref{96}, then we must simultaneously omit only the equations containing the divergence of the dual tensor from group (A) which corresponds to Maxwell's system. These equations are then consequences of group (B) and no longer belong to the system as determinative elements. Then the ``dual'' construction would no longer occur in the system. We would have ten equations for ten quantities, on the one hand the four Maxwell equations:
\begin{equation*}\label{98A}
\frac{\partial F_{i \mu}}{\partial x_{\mu}} = \alpha S_i ,
\tag{98A}
\end{equation*}
and, on the other, the six feedback equations:
\begin{equation*}\label{98B}
\frac{\partial S_i}{\partial x_k} - \frac{\partial S_k}{\partial x_i} = \alpha F_{ik} .
\tag{98B}
\end{equation*}
The divergence equation for charge is now a consequence of the equations as are the missing four Maxwellian equations.\footnote{\emph{Editorial note:} Eqs.~\eqref{98A} and~\eqref{98B} are the correct wave-equations for a massive spin~1 particle, to be rediscovered by Proca in 1936.  For more details, see section~11 in \cite{GSPON1998B}.}

   However, this system has become so non-symmetric that its immediate connection with the Dirac equations seems to be questionable, though it is also of first-order and the Schr\"odinger wave equation is obtained for each component here as well. Indeed, we arrive at the Dirac equation again if we complete our system by the second-order Maxwell equations and by the divergence equation for current --- which really hold as identities even if they do not add anything new to the system. The eight Dirac equations for the quantities $\psi$ again form a subsystem of our whole system and contain somewhat less information than our ten equations. However, here we already encounter a constraint in the $\psi$ quantities. For we can see from our assignment \eqref{55} that $\psi_1 = S_z - i S_t$ becomes a purely real quantity --- since the $G$ vector is now no longer complex but real in its space component and imaginary in its time component. The remaining $\psi$ quantities do not show a similar specialization.

   It is obvious that the question as to whether a real meaning can be attributed to the above interpretation cannot be answered until the equations are completed for the presence of an external field. Our goal did not take us so far. We raised the question as to whether the Dirac theory can be formulated by exclusively using field theoretically meaningful quantities. We have found a way which was almost unequivocally compelling and indeed led to the expected results. It resulted in an internally highly consistent covariant system which permits new prospects in some respects. Also, a certain arbitrariness --- which, as we have shown, is inherent in the Dirac transformation theory and suggests an underlying reason --- is eliminated here, since here any uncertainty is avoided in the transformations.

   Should this chosen way prove unsuccessful, it would most likely be hopeless to expect a field theoretical background for the Dirac theory. Our developments reveal little doubt that this background can lie only in the direction found here, if it exists at all.

   In formal respects, one can consider it a gain that a method has been worked out which enables the transformation properties of the Dirac quantities to be described in a very unified and transparent way.

   Berlin-Nikolassee, July 1929.

\end{document}